\documentstyle[12pt]{article}

\newcommand{\be}{\begin{equation}}
\newcommand{\ee}{\end{equation}}
\newcommand{\bn}{\begin{eqnarray}}
\newcommand{\en}{\end{eqnarray}}
\newcommand{\p}{\partial}

\def\l{\label}

\begin{document}

\begin{center}

\noindent{\large\bf
{\huge Chiral boson on a circle revisited}
}\vspace{10mm}

\noindent{
{\Large Everton M. C. Abreu} 
}\vspace{3mm}

\noindent
\begin{large}
{\it Departamento de F\'{\i}sica, Instituto de Ci\^encias Exatas, \\ Universidade Federal Rural do Rio de Janeiro,\\
BR 465-07, 23851-180, Serop\'edica, Rio de Janeiro, Brazil\\
\bigskip{\sf E-mail: evertonabreu@ufrrj.br}}\\
\end{large}
\vspace{4mm}

\today

\vspace{14mm}

\end{center}

 





\begin{abstract}
\noindent  It is a well known result that the scalar field spectrum is composed of two chiral particles (Floreanini-Jackiw particles) of opposite chiralities.  Also, that a Siegel particle spectrum is formed by a nonmover field (a Hull's noton) and a FJ particle.  We show here that, in fact, the spectrum of the chiral boson on a circle has a particle not present in its currently well known spectrum.\end{abstract}


\newpage

\section{Introduction}

The study of chiral bosons in a $D=2$ flat space has a wide interest. They
occur as basic ingredients in the formulation of string theories and in a
number of two-dimensional statistical systems, like the fractional quantum
Hall effect phenomenology.  More recently, the M-theory can be achieved
by treating the chiral sectors in a more independent way. In superior
dimensions, the six-dimensional chiral bosons belong to the supergravity and
tensor multiplets in $N=1$, $D=6$ supergravity theories. They are necessary
to complete the $N=2$, $D=6$ supermultiplet of the $M$-theory five-brane.
Finally, a ten-dimensional chiral bosons appears in $IIB$, $D=10$
supergravity.

The dual projection technique [1-6], strictly related
to canonical transformations \cite{bg}, has been presented initially in the
study of electromagnetic duality groups \cite{wotzasek}. The difference
between both concepts is that the dual projection is performed at the level
of the actions whereas a canonical transformation is performed at the
Hamiltonian level.  The most useful and
interesting point in this dual projection procedure is that it is not based
on evidently even-dimensional concepts and may be extended to the
odd-dimensional situation \cite{baw} and also in non-Abelian systems \cite{bw}. 
In this work we will show that besides
the use of dual projection in even and odd dimensions in flat space, we can
obtain interesting results in curved space.

In this work we want to study the spectrum of a chiral boson on a circle (CBC), which was described in \cite{rzt} as being formed by one FJ's particle when it is generalized on the line.  We claim here that in fact it has an additional particle, a noton.

The sequence of the paper is: in order to make this work self-consistent, in
the next section we review the $D=2$ dual projection applied to the scalar
field in a Minkowski space and to the Siegel chiral model, which
introduces the presence of the noton as in \cite{aw}.  
In section 3 we perform the dual projection of the chiral boson on a circle. 
As usual, the conclusion and perspectives are discussed in the last section.

\section{The dual projection in a flat space: a review}\label{secao2}

The separation of a scalar field into its chiral components has been
introduced by Mandelstam \cite{mandelstam} in its seminal paper on 2D
bosonization. The chiral splitting for the non-Abelian side has been studied
by Polyakov and Wiegman \cite{pw}. However, the Abelian limit of
Polyakov-Wiegman decomposition does not coincide with Mandelstam's chiral
decomposition. The chiral separation in \cite{mandelstam} is based on a
first-order theory while that of \cite{pw} is second-order. Mandelstam's
chiral decomposition scheme can be obtained as a constraint over the theory
requiring the complete separability of the original action.

The dual projection in $D=2(2p+1)$ dimensions leads to a diagonal form of
the action (hence we can also call it as a diagonalization procedure). 
Here, differently from \cite{bw}, the two pieces manifest
completely unlike features: while one piece is chiral (a FJ particle), responsible for the
dynamical sector of the theory, the other carry the algebraic component (a noton particle), as shown in \cite{aw}. We select here only two examples, among others, in the
literature where the dual projection succeed.

\subsection{The scalar field}

In this section we will make a brief analysis of the dual projection of a
scalar field in a $D=2$ Minkowski space. This result was obtained a few
years back by Tseytlin and West \cite{tw}. Performing the inverse process,
confirming this result, the recent soldering formalism \cite{solda} 
showed us that the fusion of two FJ modes gives a scalar field. 
As in this work we will consider only first-order actions, the equivalence between the dual projection and the canonical transformations becomes manifest \cite{bg}.

In a few steps we can begin with the Lagrangian density of a scalar field, 
\be  \label{1.1}
{\cal L} = {\frac{1 }{2}}\, \partial_{\mu}\,\phi\,\partial^{\mu}\,\phi 
\,=\, {\frac{1 }{2}}\,{\dot{\phi}}^2\,-\,{\frac{1 }{2}}\,{\phi^{\prime}}^2,
\ee
where the metric is $(+,-)$.  Dot and prime means respectively the time
and space derivatives. Reducing the order of the time derivative, we can use
an auxiliary field (not necessarily the canonical momentum) so that we can
write 
\begin{equation}  \label{2.11}
{\frac{1 }{2}}\,{\dot{\phi}}^2\,=\, \dot{\phi}\,p\,-\,{\frac{1 }{2}}\,p^2
\end{equation}
and substituting (\ref{2.11}) in (\ref{1.1}) the Lagrangian reads, 
$$ 
{\cal L}_1 \,=\, \dot{\phi}\,p\,-\,{\frac{1 }{2}}\,p^2 \,-\,{\frac{1 }{2}}\,{%
\phi^{\prime}}^2.
$$
Performing the canonical transformations conveniently constructed in order to
promote a complete field separation, we have, 
\begin{equation}  \label{4.1}
\phi\,=\, \rho\,+\,\sigma \qquad \mbox{and} \qquad
p\,=\,\rho^{\prime}\,-\,\sigma^{\prime}\;\;.
\end{equation}
Substituting it in ${\cal L}_1$ above, we have a new action, 
\begin{equation}
{\cal L}_2\,=\,\dot{\rho}\,{\rho^{\prime}}^2\,-\,{\rho^{\prime}}^2\,-\,\dot{%
\sigma}\,\sigma^{\prime}\,-\,{\sigma^{\prime}}^2\,\,,
\end{equation}
which shows clearly that the spectrum of a scalar field is formed by two FJ
particles of opposite chirality as we said above. For an interested reader,
the relation between two chiral particles was discussed at the Hamiltonian
level in \cite{sonnenschein}.

There are indications that a deeper understanding of such issues as string
dynamics and fractional quantum Hall effect phenomenology can be achieved by
treating the chiral sectors in a more independent way. However, coupling
chiral fields to external gauge and gravitational fields is problematic. As
we said above, it was discussed in \cite{bw}, how the coupling of chiral
(Abelian) fields to external gravitational backgrounds can be achieved by
diagonalization (dual projection) of the first-order form of a covariant
scalar action. The theory reduces then to a sum of a left and a right FJ's
actions \cite{fj}, circumventing the problems caused by the lack of manifest
Lorentz invariance.

\subsection{The Siegel chiral boson}

A noton is a nonmover field at the classical level \cite{hull}, carrying
the representation of the Siegel symmetry \cite{siegel}, that acquires
dynamics upon quantization. As another feature, it is a well known fact that
coupling chiral particles to external gravitational fields reveals the
presence of notons \cite{dgr}. At the quantum level in flat space, it was
shown \cite{aw} that its dynamics is fully responsible for the Siegel
anomaly. In other words, we can say that the importance of the inclusion of
a normalized external noton - the Hull mechanism - is to cancel the Siegel
anomaly \cite{hull} conveniently.

On the other hand, later than Siegel, Floreanini and Jackiw (FJ) introduced
an action which describes a free chiral boson in two dimensions \cite{fj},
but this action have some symmetry problems. In \cite{aw} it was shown that,
in fact both models (Siegel and FJ) for the chiral bosons are connected by a
noton particle.

The Siegel original classical Lagrangian density for a chiral
scalar field \cite{siegel} using the usual light-cone variables is, 
\be
\label{siegel}
{\cal L}_{Siegel}=\partial_+\phi\,\partial_-\phi - \lambda_{++}(\partial_-%
\phi)^2  
={\frac{1 }{2}}\,\sqrt{-g}\,g^{\alpha\beta}\partial_{\alpha}\phi\,\partial_{\beta}\,\phi
\ee
where the metric is 
$
g^{++}\,=\,0\:,\:g^{+-}\,=\,1\:,\:g^{--}\,=\,-\,2\,%
\lambda_{++}\:,
$
and we can say that the Lagrangian ${\cal L}_{Siegel}$ describes a lefton \cite{gs}.
A lefton (or a righton) is a particle which, besides to carry the dynamics
of the theory, it is liable for the symmetry of the system too. This
characterizes exactly a Siegel mode. Hence, it is different from a FJ's
particle \cite{fj} which carries only the dynamics of the system. Thereby, a
FJ particle can not be classified either as a lefton or a righton.

The symmetry content of the theory is well described by the Siegel algebra,
a truncate diffeomorphism that disappears at the quantum level. Hence ${\cal L}_{Siegel}$ is invariant under Siegel gauge symmetry which is an invariance under
the combined coordinate transformation and a Weyl rescaling of the form 
$
x^- \rightarrow \tilde{x}^-\,=\,x^-\,-\,\epsilon^- \quad \mbox{and} \quad
\delta g_{\alpha\beta} \,=\, -\,g_{\alpha\beta}\,\partial_-\,\epsilon^- .
$
The fields $\phi$ and $\lambda_{++}$ transform under these relations as
$
\delta \phi= \epsilon^-\,\partial_-\,\phi,$ and $  
\delta \lambda_{++}=
-\,\partial_+\,\epsilon^- +\epsilon^-\,\partial_-\,\lambda_{++}\,-\,
\partial_-\,\epsilon^-\,\lambda_{++}$,
and $\phi$ is invariant under the global axial transformation,
$
\phi \rightarrow \tilde{\phi}\,=\,\phi\,+\,\bar{\phi}
$.  It is easy to see that fixing the value of the multiplier as $\lambda_{++}=1$
in ${\cal L}_{Siegel}$ we can obtain the FJ model. As mentioned before, this was
considered, for a long time in the literature, the unique constraint between
these both representations of a chiral boson. The dual projection permit us
to see other intrinsic relations behind the model that can possibly (or not)
be hidden in its spectrum.

We will now begin to apply the dual projection procedure. This is done
introducing a dynamical redefinition in the phase space of the model. Using ${\cal L}_{Siegel}$ in Lorentz coordinates we can obtain the canonical momentum as 
\begin{equation}  \label{quatro}
\pi\,=\,\frac{\partial {\cal L}}{\partial \dot{\phi}}\,=\,\dot{\phi}%
\,-\,\lambda_{++}\,(\,\dot{\phi}\, -\,\phi^{\prime}\,)
\end{equation}
or, in other words 
\begin{equation}  \label{cinco}
\dot{\phi}\,=\,\frac{\pi\,-\,\lambda_{++}\,\phi^{\prime}}{1\,-\,\lambda_{++}}%
\;\;.
\end{equation}
After a little algebra, substituting (\ref{quatro}) and (\ref{cinco}) into ${\cal L}_{Siegel}$, the Lagrangian in the first-order form reads 
\begin{equation}  \label{cincoa}
{\cal L}_{Siegel}\,=\,\pi\,\dot{\phi}\,-\,\frac{{\phi^{\prime}}^2}{2}\,-\, {\frac{1 }{%
2}}\,\frac{(\,\pi\,-\,\lambda_{++}\,\phi^{\prime}\,)^2}{1\,-\,\lambda_{++}}%
\,-\, {\frac{\lambda_{++} }{2}}\,{\phi^{\prime}}^2 .
\end{equation}

As we said above, fixing the value of the multiplier as $\lambda
_{++}\rightarrow 1$ in (\ref{cincoa}) we get the FJ form. This value of $%
\lambda _{++}$ promotes a reduction of the phase space of the model to \cite%
{djt} 
$
\pi \rightarrow {\phi ^{\prime }} ,
$
and consequently the third term in (\ref{cincoa}) reduces to zero as $%
\lambda _{++}\rightarrow 1$. Therefore the dynamics of the system will be
described by a FJ action.

The above behavior in $\pi \rightarrow {\phi ^{\prime }} $ suggests the following canonical
transformations, analogous as in (\ref{4.1}):
\begin{equation}
\phi \,=\,\varphi \,+\,\sigma \qquad \mbox{and}\qquad \pi \,=\,\varphi
^{\prime }\,-\,\sigma ^{\prime }\;\;,  \label{sete}
\end{equation}
and we stress that these fields are independent as they originate from
completely different actions. After substituting (\ref{sete}) into (\ref%
{cincoa}) to perform the dual projection we find a diagonalized Lagrangian, 
\begin{equation}
{\cal L}\,=\,\varphi ^{\prime }\,\dot{\varphi}\,-\,{\varphi ^{\prime }}%
^{2}\,-\,\sigma ^{\prime }\,\dot{\sigma}\,-\,\eta _{+}\,{\sigma ^{\prime }}%
^{2} ,  \qquad \eta _{+}\,=\,\frac{1\,+\lambda _{++}}{1\,-\lambda _{++}}\;\;. 
\label{oito}
\end{equation}

The effect of dual projection procedure into the first-order Siegel theory,
equation (\ref{cincoa}), was the creation of two different internal spaces
leading to the $Z_{2}$ group of dualities (a discrete group with two
elements) \cite{wotzasek,baw} and the other is the diffeomorphism group of
transformations. Clearly we can see that the chirality/duality group and the
symmetry group are in different sectors. The first is obviously a FJ mode
and the other is a noton mode. This result is complementary to the
established knowledge, where the FJ action is interpreted as a gauge fixed
Siegel action \cite{siegel}. Under this point of view, we look at the gauge
fixing process as the condition that sets the noton field to vanish. It can
be proved \cite{aw} that this noton is totally responsible for the
symmetries, both classically and quantically.

\section{Chiral boson on the circle}

In \cite{rzt}, the CBC was shown in a generalized form which is equivalent to just one FJ's particle on the real axis.  We show in a precise way, that the CBC spectrum comprises, besides a FJ particle, a noton field.
Namely, as what we have seen above, the CBC spectrum is analogous to Siegel's.
 
The action for a CBC \cite{rzt} of length $2\pi$ is,
\be
\l{1}
{\cal L} = {1\over2}\,(\p_{\tau}\phi)^2\,-\,{1\over2}(\p_{\alpha}\phi)^2
\ee
subjected to the constraint
\be
\l{vinculo}
\p_\tau \phi \,-\, \p_\alpha \phi \approx 0\,\,.
\ee
Obviously, although the action (\ref{1}) has the same form as (\ref{1.1}), we will see that the constraint (\ref{vinculo}) will bring a new particle in CBC' spectrum.

The action above can be written as,
\be
\l{2}
{\cal L}\,=\, {1\over2} {\dot{\phi}}^2\,-\,{1\over2} {\phi^{\prime}}^2\,\,,
\ee
where 
$
{\phi'} (\alpha) = \frac{\phi(\alpha)}{d \alpha}\;\;.
$
To reduce the order of the time derivative in (\ref{2}) as in (\ref{1.1}), we can use
an auxiliary field (not necessarily the canonical momentum) so that we can
write 
\begin{equation}  
\label{dois_1}
{\frac{1 }{2}}\,{\dot{\phi}}^2\,=\, \dot{\phi}\,p\,-\,{\frac{1 }{2}}\,p^2
\end{equation}
and substituting (\ref{dois_1}) in (\ref{1}) the first-order Lagrangian reads, 
\begin{equation}  
\label{3}
{\cal L}_1 \,=\, \dot{\phi}\,p\,-\,{\frac{1 }{2}}\,p^2 \,-\,{\frac{1 }{2}}\,{%
\phi^{\prime}}^2\,\,.
\end{equation}

Constructing the canonical transformations conveniently, as in the last section, in order to promote a complete field diagonalization, we have, 
\begin{equation}  
\label{4}
\phi\,=\, \rho\,+\,\sigma \qquad \mbox{and} \qquad
p\,=\,\rho^{\prime}\,-\,\sigma^{\prime}\;\;,
\end{equation}
where $\rho'$ and $\sigma'$ are given as above.
Substituting it in ${\cal L}_1$ above, we have a new action, 
\begin{equation}
{\cal L}_2\,=\,\dot{\rho}\,{\rho^{\prime}}^2\,-\,{\rho^{\prime}}^2\,-\,\dot{%
\sigma}\,\sigma^{\prime}\,-\,{\sigma^{\prime}}^2\;\;,
\end{equation}
or,
\be
\l{5}
{\cal L}_2 \,=\,\p_{\tau} \rho \p_{\alpha} \rho\,-\,(\p_{\tau} \rho)^2 \,-\,\p_{\tau} \sigma\p_{\alpha} \sigma \,-\,(\p_{\alpha} \sigma)^2\,\,.
\ee
However, the fields in (\ref{5}) are still subjected to the constraint (\ref{vinculo}), which can be written as  
\be
p\,-\,\phi'\,=\,0
\ee
and using equation (\ref{4}) we have that
\be
\sigma'\,=\,0\,\,,
\ee
this condition is the same result coming from (\ref{oito}), which shows clearly that the spectrum of a CBC is a generalization for the circle of the model given by one FJ's particle (the field $\rho$) and a noton (the field $\sigma$).  As we said, it has the spectrum analogous to the Siegel model does.  This result is different of the usual because in this case the constraint forces the $\sigma$ field to behave like a noton.  In the other cases the noton couples to the gravitational field. 

\subsection{Dirac brackets analysis}

To discuss the Dirac constraints, as in \cite{rzt}, in a DP perspective, we have to consider periodic boundary conditions in (\ref{1}), i.e.,
$
\phi(\alpha)\,=\,\phi(\alpha\,+\,2\pi)
$, 
by introducing the variables constructed in \cite{rzt},
\bn
\l{12}
T^R(\alpha) &=& {1\over \sqrt{2}} [\pi(\alpha)\,-\,\phi'(\alpha)] \nonumber \\
T^L(\alpha) &=& {1\over \sqrt{2}} [\pi(\alpha)\,+\,\phi'(\alpha)]\;\;,
\en
where the Hamiltonian density corresponding to (\ref{1}) is 
\be
\l{13}
{\cal H} \,=\,{1\over2}\,\pi^2(\alpha)\,+\,{1\over2}\, {\phi'}^2(\alpha)\;\;,
\ee
and defining the Dirac brackets as functions of the corresponding components of $T^R (\alpha)$ and $T^L (\alpha)$ (further details can be found in \cite{rzt}).

It can also be seen directly from (\ref{4}) that we can write 
\bn
\l{18}
\phi' = \rho'\,+\, \sigma' \qquad \mbox{and} \qquad
p = \rho'\,-\, \sigma'\;\;,
\en
and using (\ref{12}) it is easy to see that  we can write the DP auxiliary fields $\rho$ and $\sigma$ submitted to the periodic boundary conditions as 
\bn
\rho(\alpha) &=& {1\over\sqrt{2}} \int T^R (\alpha')\, d\alpha' \nonumber \\
\sigma(\alpha) &=& {1\over\sqrt{2}} \int T^L (\alpha')\, d\alpha' \;\;,
\en
which permit us to talk about canonical quantization.  But this issue is beyond the aim of this letter.  From (\ref{18}) we believe that it can be demonstrated a possible connection between the DP formalism and the introduction in \cite{rzt} of the variables $T^R(\alpha)$ and $T^L(\alpha)$ when we compare equations (\ref{12}) and (\ref{18}).

\section{Conclusions}

The dual projection procedure is a technique that helps one to analyze the spectrum
of first-order systems.  It was used recently to study the notion of
duality and self-duality creating an internal space of potentials \cite{baw}.

Since the analysis was always effected for first-order systems, an
equivalence between the Lagrangian and Hamiltonian approaches permitted us
to use the concept of canonical transformations. In other words we can say
that the dual projection demanded a change of variables which was, in the
phase space, a canonical transformation.

We performed the analysis of a chiral boson on a circle and the result is different from the one obtained in \cite{rzt}.  Here we show that the spectrum is analogous to the Siegel one, i.e., a FJ's particle and a noton.  The presence of the noton is determined by the constraint acting on the fields, whereas in general the noton appears naturally coupled to the gravitational field.

We show here that any conclusion about the spectrum of a theory turns out to be premature and thereby must be preceded by a detailed (for example) dual projection analysis.  In this way we can determine precisely if the particles that comprise the spectrum can be either notons or chiral fields coupled to the gravitational background even if the final results are the same.
 
\bigskip

\section{Acknowledgments}

The author would like to thank Dr. Marco Moriconi for valuable discussions and Funda\c{c}\~ao de Amparo \`a Pesquisa do Estado do Rio de Janeiro (FAPERJ) for partial financial support.




\end{document}